\documentclass[12pt,preprint]{aastex}

\begin{document}

\title{Hadronic high-energy gamma-ray emission from the microquasar LS I $+61 \, 303$}

\author{Gustavo E. Romero\altaffilmark{1}}
\affil{Instituto Argentino de Radioastronom\'{\i}a (IAR), C.C.\ 5, 1894 Villa Elisa,
Argentina.} 
\email{romero@irma.iar.unlp.edu.ar}
\author{Hugo R. Christiansen}
\affil{State Univesity of Ceara, Physics Dept., Av. Paranjana 1700, 60740-000 Fortaleza - CE, Brazil}
\email{hugo@uece.br}
\and
\author{Mariana Orellana\altaffilmark{2}}
\affil{Instituto Argentino de Radioastronom\'{\i}a (IAR), C.C.\ 5, 1894 Villa Elisa,
Argentina.} 
\email{morellana@irma.iar.unlp.edu.ar} 

\altaffiltext{1}{Member of CONICET, Argentina}
\altaffiltext{2}{Fellow of CONICET, Argentina}

\begin{abstract}
We present a hadronic model for gamma-ray production in the microquasar LS I $+61 \, 303$. The system is formed by a neutron star that accretes matter from the dense and slow equatorial wind of the Be primary star. We calculate the gamma-ray emission originated in $pp$ interactions between relativistic protons in the jet and cold protons from the wind. After taking into account opacity effects on the gamma-rays introduced by the different photons fields, we present high-energy spectral predictions that can be tested with the new generation Cherenkov telescope MAGIC. 
\end{abstract}

\section{Introduction}
LS I $+61 \, 303$ is a Be/X-ray binary that presents unusually strong and variable radio emission (Gregory \& Taylor 1978). The X-ray emission is weaker than in other objects of the same class (e.g. Greiner \& Rau 2001) and shows a modulation with the radio period (Paredes et al. 1997). The most recent determination of the orbital parameters (Casares et al. 2005) indicates that the eccentricity of the system is $0.72\pm0.15$ and that the orbital inclination is $\sim 30^{\circ}\pm 20^{\circ}$. The best determination of the orbital period ($P=26.4960\pm0.0028$) comes from radio data (Gregory 2002). The primary star is a B0 V with a dense equatorial wind. Its distance is $\sim 2$ kpc. The X-ray/radio ourtbursts are triggered 2.5-4 days after the periastron passage of the compact object, usually thought to be a neutron star. These outbursts can last until well beyond the apastron passage.

Recently, Massi et al. (2001) have detected the existence of relativistic radio jets in LS I $+61 \, 303$, which makes of it a member of the microquasar class. Microquasars are thought to be potential gamma-ray sources (Paredes et al. 2000, Kaufman Bernad\'o et al. 2002, Bosch-Ramon et al. 2005a) and, in fact, LS I $+61 \, 303$ has long been associated with a gamma-ray source. First with the COS-B source CG135+01, and later on with 3EG J0241+6103 (Gregory \& Taylor 1978, Kniffen et al 1997). The gamma-ray emission is clearly variable (Tavani et al. 1998) and has been recently shown that the peak of the gamma-ray lightcurve is consistent with the periastron passage (Massi 2004), contrary to what happens with the radio/X-ray emission, which peaks {\sl after} the passage.

The matter content of microquasars jets is unknown, although in the case of SS 433 iron X-ray line observations have proved the presence of ions in the jets (Kotani et al. 1994, 1996; Migliari et al. 2002). In the present paper we will assume that relativistic protons are part of the content of the observed jets in LS I $+61 \, 303$ and we will develop a simple model for the high-energy gamma-ray production in this system, with specific predictions for Cherenkov telescopes like MAGIC. We emphasize that our model is not opposed, but rather complementary to pure leptonic models as those presented by Bosch-Ramon \& Paredes (2004) and Bosch-Ramon et al. (2005a), since the leptonic contribution might dominate at lower gamma-ray energies and after the periastron passage. In the next section we will describe the basic features of the model, and then we will present the calculations and results.

\section{General picture}

A hadronic model for the gamma-ray emission in microquasars with early-type companions has been already developed by  Romero et al. (2003). This model, however, is limited to the simple case of a massive star with a spherically symmetric wind and a compact object in a circular orbit. Here we will consider a B-type primary with a wind that forms a circumstellar outflowing disk of density $\rho_{\rm w}(r)=\rho_0({r}/{R_*})^{-n}$ (Gregory \& Neish, 2002). The continuity equation implies a wind velocity of the type $v_{\rm w} = v_0 (r/R_*)^{n-2}$. We will consider that the wind remains mainly near to the equatorial plane, confined in a disk with half-opening angle $\phi=15^\circ$, with $n=3.2$, $\rho_0=10^{-11}$ g cm$^{-3}$, and $v_0=5$ km s$^{-1}$ (Mart\'{\i} and Paredes 1995).

The modeled properties of the system will be expressed in terms of the orbital phase $\psi$ ($\psi=0.23$ at the periastron passage according to the latest determination by Casares et al. 2005) which is related to the separation between the stars by $r(\psi)=a (1-e^2)/[1-e \cos (2\pi (\psi+0.73))]$, where $a$ is the semi-major axis of the orbit and $e$ the eccentricity. 
The wind accretion rate onto the compact object of mass $M_{\rm c}$ can be estimated as:
\begin{equation}
\dot{M}_{\rm c}=\frac{4 \pi (G\,M_{\rm c})^2 \rho_w(r)}{v_{\rm rel}^3},
\label{macc}
\end{equation}
where $v_{\rm rel}$ is the relative velocity between the neutron star (moving in a Keplerian orbit) and the circumstellar wind, assumed to be flowing radially on the equatorial plane.

Following the basic assumption of the jet-disk symbiosis model (Falcke \& Biermann 1995) we will assume that the accretion rate is coupled to the kinetic jet power by:
\begin{equation}
Q_{\rm j}=q_{\rm j} \dot{M}_{\rm c} c^2,
\end{equation}
where $q_{\rm j}\sim 0.1$ is the coupling constant. Most of this power will consist of cool protons that are ejected with a macroscopic Lorentz factor $\Gamma\sim 1.25$ (Massi et al. 2001). Only a small fraction 
$q^{\rm rel}_{\rm j}\sim 10^{-3}$ is in the form of relativistic hadrons. The relativistic jet is confined by the pressure of the cold particles ($P_{\rm cold}>P_{\rm rel}$), which expand laterally at the local sound speed.

The jet axis, $z$, will be assumed normal to the orbital plane. The jet will be conical, with a radius $R_{\rm j}(z)= z (R_0/z_0)$, where $z_0$ is the injection point and $R_0$ is the initial radius of the jet. We will adopt  $z_0= 10^7$ cm and $R_0=z_0/10$ as reasonable values (see Romero et al. 2003 and Bosch-Ramon et al. 2005a, who deals with similar jets for additional details). The relativistic proton spectrum will be a power law $N'_{p}(E'_{p})= K_p\;
{E'}_{p}^{-\alpha}$, valid for $ {E'_{p}}^{\rm min}\leq E'_{p}
\leq  {E'_{p}}^{\rm max}$ (in the jet frame). The
corresponding relativistic proton flux will be $J'_{p}( E'_{p})= (c/4\pi)
N'_{p}(E'_{p})$. 
Since the jet expands in a conical way, the
proton flux evolves with $z$ as:
\begin{equation}
J'_p(E'_p)=\frac{c}{4 \pi} K_0
\left(\frac{z_0}{z}\right)^2 {E'_p}^{-\alpha},
\label{Jp}
\end{equation}
where it is implicit the assumption of the conservation of the number of particles (see Ghisellini et al. 1985), and a prime refers to the jet frame. Using relativistic
invariants, it can be shown that the proton flux, in the lab (observer) frame, becomes (e.g. Purmohammad \& Samimi 2001):

\begin{equation}
J_p(E_p,\theta)=\frac{c K_0}{4 \pi} \left(\frac{z_0}{z}\right)^
2 \frac{\Gamma^{-\alpha+1} \left(E_p-\beta_{\rm b}
\sqrt{E_p^2-m_p^2c^4} \cos \theta\right)^{-\alpha}}{\left[\sin ^2
\theta + \Gamma^2 \left( \cos \theta - \frac{\beta_{\rm b}
E_p}{\sqrt{E_p^2-m_p^2 c^4}}\right)^2\right]^{1/2}}. \label{Jp_lab}
\end{equation}
In this expression, $\Gamma$ is the jet Lorentz factor, $\theta$ is the angle
subtended by the proton velocity direction (which will be roughly the same as that of the emerging photon) and the jet axis (notice that then $\theta \approx \theta_{\rm obs}$), and $\beta_{\rm b}$ is the bulk velocity in units of $c$. We will make all calculations in the lab frame, where the cross sections for $pp$ interactions have suitable parametrizations. 

The number density ${n_0}'$ of particles flowing in the jet at $R_0$, and the normalization constant $K_0$ can be determined as in Romero et al. (2003). In the numerical calculations of the next section we have considered ${E'}_p^{\rm max}=100$ TeV, ${E'}_p^{\rm min}=1$ GeV, $\Gamma=1.25$, and, $\alpha=2.2$ (see the list of the assumed parameters in Table 1). The assumed maximum energy is consistent with the jet size and shock acceleration with an efficiency $\sim0.01-0.1$.

The matter from the wind can penetrate the jet from the side, diffusing into it as long as the particle gyro-radius is smaller than the radius of the jet. This imposes a constraint onto the value of the magnetic field in the jet: $B_{\rm jet}\geq E_k/(e R_0)$, where $E_k= m_p\,v_{\rm rel}^2/2$. For the periastron passage ($E_k$ maximum) results $B_{\rm jet}\geq 2.8\,10^{-6}$ G, which is surely satisfied. However, some effects, like shock formation on the boundary layers, could prevent some particles from entering into the jet. Given our ignorance of the microphysics involved, we adopt a parameter $f_{\rm p}$ that takes into account particle rejection from the boundary in a phenomenological way. In a conservative approach, we will adopt $f_{\rm p}\sim 0.1$ .  

Some of the particles entering the jet flow would be immediately accelerated to the jet velocity (by Coloumb interactions or wave-particle interactions). As a consequence, the jet should be slowed down during its motion through the equatorial wind. However, it is a fact that the jet survives this interaction since it is seen at radio wavelengths far beyond the wind region, up to distances of $\sim400$ AU (Massi et al. 2001, 2004). Since the bulk velocity seems not to be very high (Massi et al. 2001) and hence its change does not affect seriously the calculations of the gamma-ray emissivity, we will neglect, in what follows, the effects of a macroscopic deceleration. The reader interested in the case of the hadronic gamma-ray emission of a jet slowed down to rest by the effects of the wind and the resulting standing shock wave as the major source of radiation is referred to the recent treatment presented by Romero \& Orellana (2005). 

\begin{figure}%[h]
\epsscale{.80}
%\centering 
\plotone{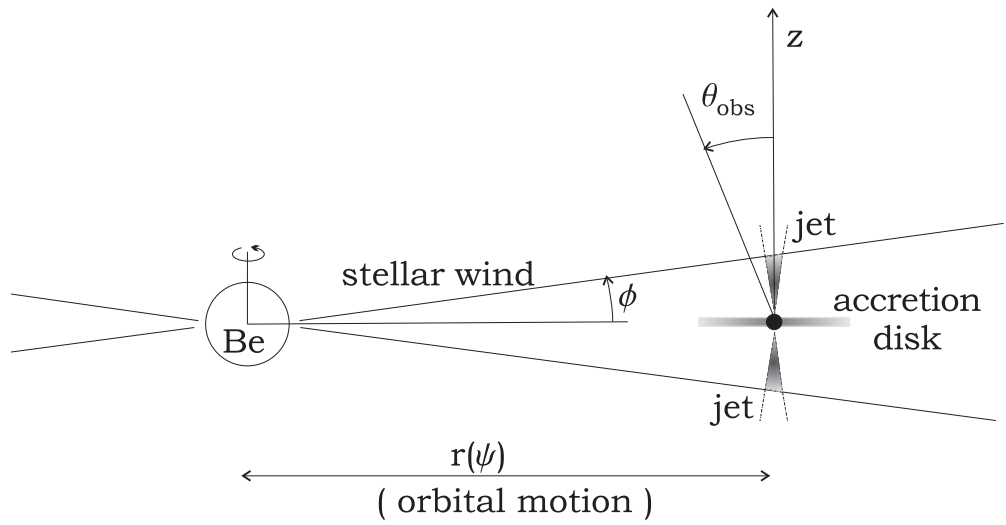} \caption{Sketch of the situation described in the paper (not drawn to scale).}\label{esquema}
\end{figure}

\begin{figure}%[h]
%\centering 
\epsscale{.60}
\plotone{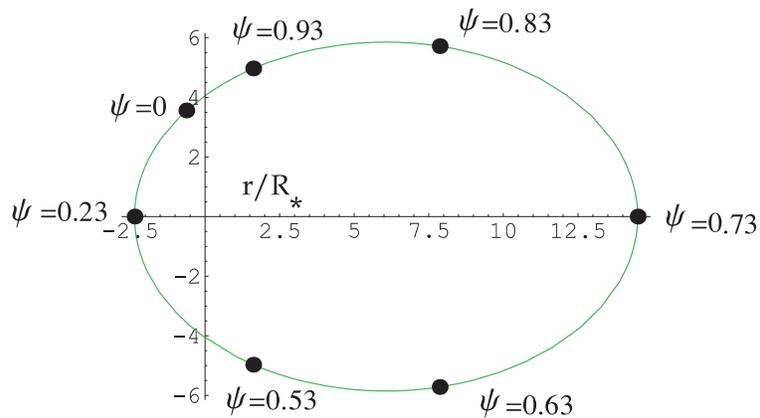} \caption{Orbit of LS I $+61 \, 303$ drawn to scale illustrating the geometry used in the calculations. %Notice that the orbital phase has been re-defined in such a way that the periastron occurs at $\psi=0.5$. The historical phase is such that 
The periastron passage occurs at 0.23 (Casares et al. 2005).}\label{orbita}
\end{figure}

In Figure \ref{esquema} we show a sketch of the general situation and in Figure \ref{orbita} we show the orbit of the system and the corresponding phases.

\section{Gamma-ray emission}

Relativistic protons in the jet will interact with target protons in the wind through the reaction channel $p+p\rightarrow p+p+ \xi_{\pi^{0}} \pi^{0} + \xi_{\pi^{\pm}} (\pi^{+} + \pi^{-})$, where $\xi_{\pi}$ is the corresponding multiplicity. Then pion decay chains will lead to gamma-ray and neutrino emission.
The differential gamma-ray emissivity from $\pi^0$-decays can be expressed as (e.g. Aharonian \& Atoyan 1996):
\begin{equation}
q_{\gamma}(E_{\gamma},\theta)= 4 \pi \sigma_{pp}(E_p)
\frac{2Z^{(\alpha)}_{p\rightarrow\pi^0}}{\alpha}\;J_p(E_{\gamma},\theta)
\,\eta_{\rm A}, \label{q} 
\end{equation} 
where $Z^{(\alpha)}_{p\rightarrow\pi^0}$ is the
so-called spectrum-weighted moment of the inclusive cross-section (see, for
instance, Gaisser 1990). $J_p(E_{\gamma})$ is the proton flux distribution (\ref{Jp_lab}) evaluated at $E=E_{\gamma}$. The cross section $\sigma_{pp}(E_p)$
for inelastic $p-p$ interactions at energy $E_p\approx 6 \xi_{\pi^{0}}
E_{\gamma}/K$, where $K\sim0.5$ is the inelasticity coefficient and $\xi_{\pi^{0}}=1.1 (E_p/\rm GeV)^{1/4}$, can be represented for $E_{p}\geq 1$ GeV by
$$\sigma_{pp}(E_p)\approx 30 \times [0.95 + 0.06 \log \;(E_p/{\rm
GeV})]\;\; (\rm mb).$$  
Finally, the parameter $\eta_{\rm A}$ takes into account the contribution from different
nuclei in the wind and in the jet (for standard composition of
cosmic rays and interstellar medium  $\eta_{\rm A}\sim1.4$).

% Notice that
%$q_{\gamma}$ is expressed in ph s$^{-1}$ sr$^{-1}$ erg$^{-1}$ when we adopt
%CGS units.

The spectral energy distribution is:
\begin{equation}
L_{\gamma}(E_{\gamma},\theta)= E_\gamma^2 \int_V n(\vec{r'})\,q_{\gamma}(E_\gamma,\theta)\,
 d^3\vec{r'}, \label{Lum}
\end{equation}
where $V$ is the interaction volume between the jet and the circumstellar disk. The particle density of the wind that penetrates the jet is $n(r)\approx f_{\rm p} \rho_w(r)/m_p$.

In our calculations, we adopt a viewing angle of $\theta = 30^\circ$ in accordance with the average value given by Casares et al. (2005). In Figure \ref{Lum} we show a 3-D plot that shows the evolution of the gamma-ray spectral energy distribution as a function of the orbital phase. Other two plots in this figure show cuts at both the periastron and apastron, and the luminosity evolution with the orbital phase at 100 GeV. In both cases we show the unabsorbed (dashed lines) and the absorbed (continuum lines) curves. This absorption is discussed in the next section.

At the periastron passage the unattenuated luminosity is $\sim 10^{33}$ erg s$^{-1}$. We can make a simple order-of-magnitude estimate of this value. The accretion rate at the periastron is $\sim 3\times 10^{17}$ g s$^{-1}$. This means that the total power in relativistic protons should be $Q^{\rm rel}_{\rm j}=10^{-3} \dot{M}_{\rm c} c^2 \sim 2.8 \times 10^{35}$ erg s$^{-1}$. The density of the stellar wind at the injection point of the jet is $n\sim 4\times 10^{11}$ cm$^{-2}$ and the cross section for protons of $E_p\sim 1$ TeV, $\sigma_{pp}\sim 34$ mb. Hence, the mean free path of the protons results $\lambda_{pp}\sim 8.3 \times 10^{13}$ cm. The thickness of the region of the disk traversed by the jet is $\Delta z \sim r_{\rm perias} \tan 15^{\circ}\sim 4.4 \times 10^{11}$ cm. Consequently, we can approximate the gamma-ray luminosity by:

\begin{equation}
L_{\gamma}=2 f_{\pi} Q^{\rm rel}_{\rm j} \left(1-e^{-\Delta z/\lambda_{pp}}\right),\label{L}
\end{equation}
where $f_{\pi}\sim 0.2$ is the fraction of the energy of the leading proton that goes into neutral pions and hence into gamma-rays. With a simple substitution into Eq. (\ref{L}) we get $L_{\gamma}\sim 6.6 \times 10^{32}$ erg s$^{-1}$, in good agreement with the detailed numerical calculations presented in Fig. \ref{Lum}.

\section{Opacity}

The optical depth for a photon with energy $E_{\gamma}$, which in this case depends upon the direction observed, can be estimated as
 
\begin{equation}
\tau(\rho,\,E_{\gamma})=\int_{E_{\rm
min}(E_{\gamma})}^\infty\int_{\rho}^\infty\,n_{\rm ph}(E_{\rm
ph},\rho')\sigma_{e^-e^+}(E_{\rm ph},E_{\gamma})d\rho'\,dE_{\rm ph}, \label{tauXg}
\end{equation}
where $E_{\rm ph}$ is the energy of the ambient photons, $n_{\rm
ph}(E_{\rm ph},\rho)$ is their density at a distance $\rho$ from the neutron star, and $\sigma_{e^-e^+}(E_{\rm ph},E_{\gamma})$ is the photon-photon pair creation cross section given by:
\begin{equation}
\sigma_{e^+e^-}(E_{\rm ph}, \;E_{\gamma})=\frac{\pi
r_0^2}{2}(1-\xi^2)\left[2\xi(\xi^2-
2)+(3-\xi^4)\ln\left(\frac{1+\xi}{1-\xi}\right) \right],
\end{equation}
where $r_0$ is the classical radius of the electron and 
\begin{equation}
\xi=\left[1-\frac{(m_e c^2)^2}{E_{\rm ph} E_{\gamma}}\right]^{1/2}.
\end{equation}
In Eq. (\ref{tauXg}), $E_{\rm min}$ is the threshold energy for pair creation in the ambient photon field. This field can be considered as formed by two components, one from the Be star and the other from the hot accreting matter impacting onto the neutron star: $n_{\rm ph}=n_{\rm ph,1}+ n_{\rm ph,2}$. Here,
\begin{equation}
n_{\rm ph,1}(E_{\rm ph},\rho)= \left(\frac{\pi B(E_{\rm
ph})}{hc\,E_{\rm ph}}\right)\frac{R_\star ^2}{\rho^2+r^2-2\rho r \sin\theta}\;,
\end{equation}
is the black body emission from the star, with
\begin{equation}
B(E_{\rm ph})= \frac{2 E_{\rm ph}^3}{(hc)^2\,(e^{E_{\rm ph}/kT_{\rm eff}}-1)}
\end{equation}
and $T_{\rm eff}=22500$ K (Mart\'{\i} \& Paredes 1995). The separation $r$ between the stars is again variable with the phase angle $\psi$.

The emission from the heated matter can be approximated by a Bremsstrahlung spectrum:
\begin{equation}
n_{\rm ph,2}(E_{\rm ph},\rho)=\frac{L_X\,E_{\rm
ph}^{-2}}{4\pi c\,\rho^2\,e^{E_{\rm ph}/E_{\rm cut-off}}}\mbox{  for $E_{\rm ph}\geq 1$ keV},
\end{equation}
where $L_X$ is the total luminosity in hard X-rays and $E_{\rm cut-off}\sim 100$ keV. The photon index of the hard X-rays is taken to be within the range published by Greiner \& Rau (2001), which was observationally determined.  $L_X$ is also constrained by observations, being $L_X\sim 10^{34}$ erg s$^{-1}$ (Paredes et al. 1997). Notice that no bump due to a putative accretion disk has been observed at X-rays, so we neglect this contribution.

\begin{figure}%[h]
%\centering 
\epsscale{.60}
\plotone{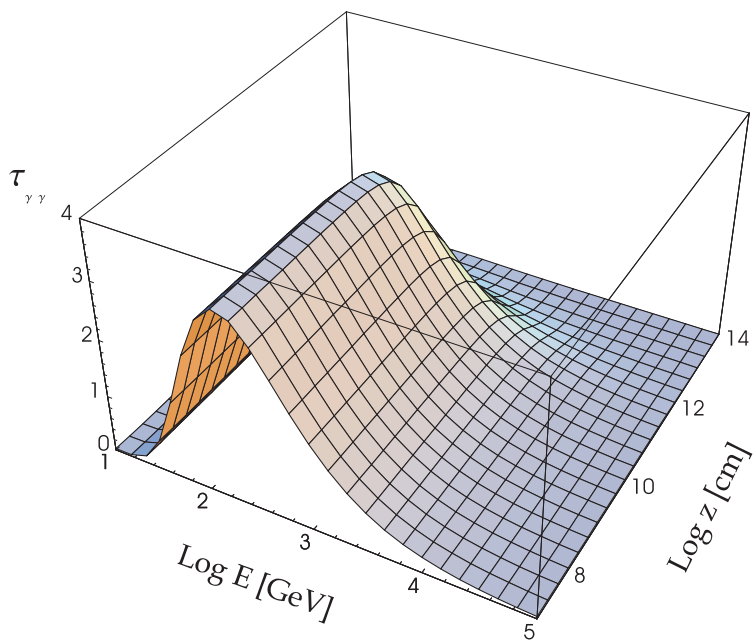} \caption{Opacity of the ambient photon fields to gamma-ray propagation. The figure corresponds to a viewing angle $\theta_{\rm obs}=30^\circ$ and $\psi=0.23$ (periastron).}\label{tau}
\end{figure}

As an example, Figure \ref{tau} shows the dependence of the optical depth $\tau$ with the energy of the $\gamma$-rays and its variation along the $z$ axis for the observer at $\theta_{\rm obs}=30^\circ$. From detailed versions of this plot, we find that for photons of $E_\gamma=100$ GeV significant absorption occurs mostly between $\psi=0.1$ and $\psi=0.5$. The optical depth remains well below the unity along the whole orbit for photons of energies $E_\gamma\la 30$ GeV and $E_\gamma\ga 2$ TeV.

\section{Secondary electron-positron pairs and synchrotron emission}

Secondary pairs are produced by the decays of charged pions and muons, as well as by photon-photon interactions. The main reactions that lead to charged pions are:
\begin{eqnarray}
p+p &\rightarrow& p+p+ \xi_{\pi^{0}} \pi^{0} + \xi_{\pi^{\pm}} (\pi^{+} + \pi^{-}) \label{pi0}\\
p+p &\rightarrow& p+ n+\pi^+ +X \label{n}\\
p+p &\rightarrow& 2n + 2\pi^+ +X \label{2n}
\end{eqnarray} 
where $n$ is a neutron, $X$ stands for anything (neutral) else, and the charged pion multiplicity is $\xi_{\pi^{\pm}}\approx 2 (E_p/\rm GeV)^{1/4}$. The neutrons have a proper lifetime of $886\pm1$ s and since they move at ultrarelativistic speed can escape from the source, decaying at considerable distances (Eichler \& Wiita 1978). On the contrary, pions decay into the jet trough $\pi^{\pm}\rightarrow\mu^{\pm}+\nu$ and $\mu^{\pm}\rightarrow e^{\pm}+\nu+\overline{\nu}$. For an injection proton spectrum given by Eq. (\ref{Jp}) with $\alpha=2.2$, we have that the pion spectrum (in the jet's system) will be a power-law $J'_{\pi^{\pm}}(E'_{\pi^{\pm}})=K_{\pi^{\pm}}\; {E'}_{\pi^{\pm}}^{-\alpha_{\pi}}$, with $\alpha_{\pi}\sim 2.3$. The electron-positron distribution mimics this power law (Ginzburg \& Syrovatskii 1964, Dermer 1986):
\begin{equation}
{J'}_{e^{\pm}}({E'}_{e^{\pm}})= K_{\pi\rightarrow e^{\pm}} {E'}_{e^{\pm}}^{-\alpha_{\pm}},    
\end{equation}
with
\begin{equation}
 K_{\pi\rightarrow e^{\pm}}=\left(\frac{m_{\mu}}{m_{e}}\right)^{\alpha_{\pm}-1} \frac{2(\alpha_{\pm}+5)}{\alpha_{\pm}(\alpha_{\pm}+2)(\alpha_{\pm}+3)} \; K_{\pi^{\pm}},    
\end{equation}
and $\alpha_{\pm}=\alpha_{\pi}$. 

The energy density of pion-generated pairs along the jet at the periastron passage can be calculated as:
\begin{equation}
w_{\pi\rightarrow e^{\pm}}= \int (4\pi/c) {E'}_{e^{\pm}} {J'}_{e^{\pm}}({E'}_{e^{\pm}}) d{E'}_{e^{\pm}},    
\end{equation}
where ${J'}_{e^{\pm}}({E'}_{e^{\pm}})$ takes into account all the contributions from $z_0$ to $z_{\rm max}$. Integrating we get $w_{\pi\rightarrow e^{\pm}}\approx 3 \times 10^{9}$ erg cm$^{-3}$.

We can compare the energy density of pairs from the charged pion decays with that of the pairs produced by direct gamma-ray absorption. The total luminosity of these pairs is:
\begin{equation}
L_{e^{\pm}}= L_{\gamma}^{0}(1-e^{-\tau}).    
\end{equation}
Then, using the opacity calculated in the previous section, the pair energy density results 
\begin{equation}
w_{\gamma\gamma\rightarrow e^{\pm}}\sim \frac{L_{e^{\pm}}}{4\pi R_0^2 c}.    
\end{equation}
At the periastron passage, we get $w_{\gamma\gamma\rightarrow e^{\pm}}\approx 3.7 \times 10^{9}$ erg cm$^{-3}$. Hence, the pair injection from the photon-photon annihilation is similar to that of pion decay. In what follows we will evaluate the spectrum of these particles using the approximation derived by Aharonian et al. (1983), which is in excellent agreement with the more detailed calculations (exact to 2nd order QED) presented by B$\ddot{\rm o}$ttcher \& Schlickeiser (1997).  

The differential pair injection rate is given by (B$\ddot{\rm o}$ttcher \& Schlickeiser 1997):

\begin{eqnarray}
\dot{n}_{e^{\pm}}(\gamma)&=& \frac{3}{32}c \sigma_{\rm _T} \int^{\infty}_{\gamma} d\epsilon_{\gamma} \frac{N_{\gamma}(\epsilon_{\gamma})}{\epsilon^{3}_{\gamma}} \int^{\infty}_{\frac{\epsilon_{\gamma}}{4\gamma(\epsilon_{\gamma}-\gamma)}} d\epsilon_{\rm ph} \frac{n_{\rm ph}(\epsilon_{\rm ph})}{\epsilon^{2}_{\rm ph}} \times  \nonumber \\
& & [\;\; \frac{4\epsilon^{2}_{\gamma}}{\gamma (\epsilon_{\gamma}-\gamma)} \ln \left(\frac{4 \epsilon_{\rm ph} \gamma (\epsilon_{\gamma}-\gamma)}{\epsilon_{\gamma}} \right) -8 \epsilon_{\gamma}\epsilon_{\rm ph}+ \frac{2(2\epsilon_{\gamma}\epsilon_{\rm ph}-1)\epsilon^{2}_{\gamma}}{\gamma(\epsilon_{\gamma}-\gamma)} -  \nonumber \\
&&  \left( 1- \frac{1}{\epsilon_{\gamma}\epsilon_{\rm ph}} \right) \frac{\epsilon^{4}_{\gamma}}{\gamma^{2}(\epsilon_{\gamma}-\gamma)^{2}}\;\;], 
\end{eqnarray}
where $\gamma=E_{e^{\pm}}/m_{\rm e} c^2$, $\epsilon_{\gamma}=E_{\gamma}/m_{\rm e} c^2$, and $\epsilon_{\rm ph}=E_{\rm ph}/m_{\rm e} c^2$. A numerical integration yields a pair spectrum that can be well fitted by a power law $N_{e^{\pm}}\propto E^{-1.9}_{e^{\pm}}$. The proportionality constant $K_{\gamma\gamma\rightarrow e^{\pm}}$ can be obtained from the absorbed gamma-ray luminosity. 
 
The presence of a magnetic field in the jet will imply that all these secondary pairs will produce synchrotron emission. Following Bosch-Ramon et al. (2005b) we assume that the magnetic field is entangled to cold protons in such a way it has random directions and hence the synchrotron emission is isotropic in the jet's frame. To calculate the synchrotron luminosity we estimate the specific emission ($j_{\epsilon}(z)$) and absorption ($k_{\epsilon}(z)$) coefficients from the secondary particle distribution (see Pacholczyk 1970 for the detailed formulae), in such a way that:

\begin{equation}
\frac{dL_{\epsilon}(z)}{dz}=2\pi R_{\rm j}
\frac{j_{\epsilon}(z)}{k_{\epsilon}(z)}\times
[1-\exp(-l_{\rm j}k_{\epsilon}(z))],
\label{eq:syncem}
\end{equation}
where to simplify the notation we are not using now primes to indicate that the calculation is in the jet's frame. In Eq. (\ref{eq:syncem}) $l_{\rm j}\sim R_{\rm j}$ is the typical size of the synchrotron emitting plasma and $\epsilon$ is the photon energy in units of $m_{\rm e}c^2$. Integrating over the jet length we get the spectral energy distribution as:
\begin{equation}
L^{\rm obs}_{\rm syn}=\epsilon  \int^{z_{\rm max}}_{z_0}
 \delta^2 \frac{dL_{\epsilon}}{dz}dz
\label{eq:L},
\end{equation}
where $\delta$ is the Doppler boosting factor defined as:
\begin{equation}
\delta=\frac{1}{\Gamma (1-\beta_{\rm b}\cos\theta_{\rm obs})}. 
\label{eq:dopb}
\end{equation}

\begin{figure}%[h]
\epsscale{.80}
%\centering 
\plotone{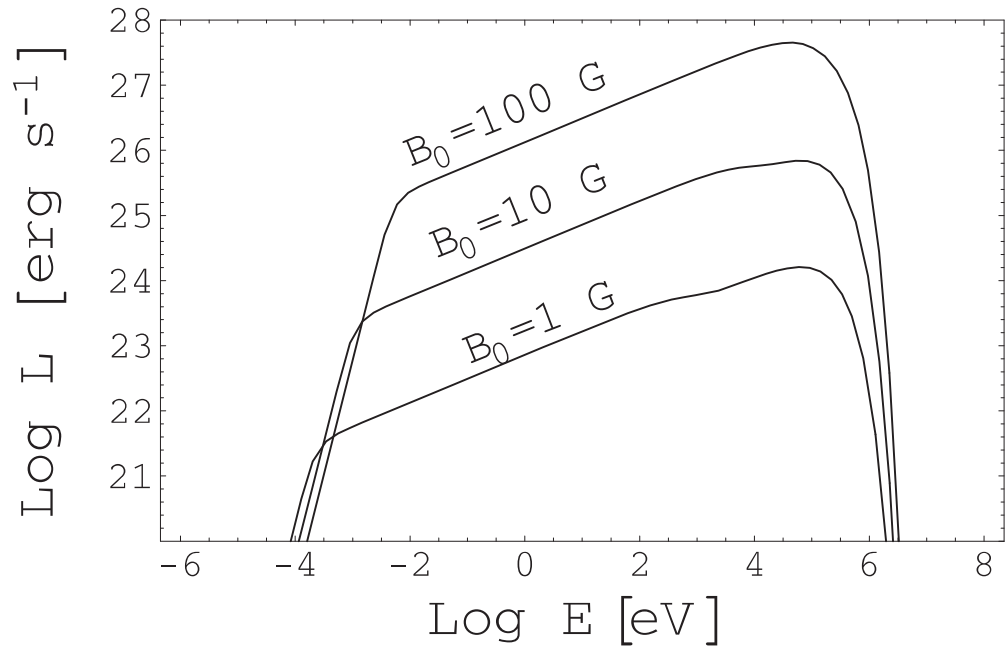} \caption{Synchrotron emission from secondary pairs.}\label{f5}
\end{figure}

To calculate the specific emission $j_{\epsilon}(z)$ we adopt different values of the magnetic field at $z_0$: $B_0=1$, 10, and 100 Gauss (Bosch-Ramon \& Paredes 2004). 
%The magnetic field evolves with $z$ as:
%\begin{equation}
%B(z)=B_0\left(\frac{R_0}{R}\right)=B_0\left(\frac{z_0}{z}\right).
%\label{eq:Bevol}
%\end{equation} 
In Figure \ref{f5} we show the spectral energy distribution of the synchrotron radiation of all secondary pairs for the 3 different values of $B_0$. The radio emission is quite negligible in comparison to the observed values, which at the minimum imply a luminosity of $\sim 10^{31}$ erg s$^{-1}$ (e.g. Rib\'o et al. 2005).

\section{Discussion}

 The predicted gamma-ray luminosity is clearly at its maximum during the periastron passage, when the neutron star travels through the densest parts of the wind. This is in accordance with the fact noticed by Massi (2004) that the peaks of the EGRET flux are coincident with the periastron and not with the radio maxima. The radio outbursts are the result of particle injection in the jet that occurs after some relaxation time from the periastron passage, when the accretion rate is increased (Paredes et al. 1991). Any purely leptonic model for the gamma-ray emission would have to explain why the radio and gamma-ray peaks are not observed in similar orbital phases.
 
Other specific feature of the gamma-ray emission predicted by our model is the presence of a local, secondary maximum at $\psi\sim 0.65$ when the accretion rate, given by (\ref{macc}), has also a local maximum due to the fact that the wind velocity is roughly parallel to the neutron star orbital velocity, hence reducing $v_{\rm rel}$ and increasing $\dot{M}_{\rm c}$, as noticed by Mart\'{\i} \& Paredes (1995). 
 
The effects of the opacity of the ambient photon fields to gamma-ray propagation produces a ``valley'' in the spectral energy distribution, between a few tens of GeV and a few TeV, with a local minimum at around 100 GeV, during the periastron passage. The predicted luminosity is within the detection possibilities of an instrument like MAGIC, which, integrating over several periastron passages, could build up a SED which can be compared with that presented in Fig. \ref{Lum}. Upper limits obtained with the Whipple telescope (Hall et al. 2003, Fegan et al. 2005) are indicated in the figure. The source is too weak for the sensitivity of this instrument according to our model.

\section{Concluding remarks}

We have presented a hadronic model for the high-energy gamma-ray production in the microquasar LS I $+61 \, 303$. The model is based on the interaction of a mildly relativistic jet with a small content of relativistic hadrons  with the dense equatorial disk of the companion B0 V star. Gamma-rays are the result of the decay of neutral pions produced by $pp$ collisions. Charged pion decay will lead to neutrino production, that will be discussed elsewhere. The model takes into account the opacity of the ambient photon fields to the propagation of the gamma-rays. The predictions include a peak of gamma-ray flux in the periastron passage, with a secondary maximum at phase $\psi\sim 0.65$. The spectral energy distribution presents a minimum around 100 GeV due to absorption. The spectral features should be detectable by an instrument like MAGIC through exposures $\sim 50$ hr, integrated along different periastron passages.     

\section*{Acknowledgments}

We thank J.M. Paredes and V. Bosch-Ramon for careful readings of the manuscript and comments. The latter gave us useful support on calculations for the secondary emission. We also thank constructive suggestions by an anonymous referee. This work has been supported by the Argentinian agencies CONICET and ANPCyT (PICT 03-13291). HRC thanks support from FUNCAP and CNPq (Brazil).

\clearpage
%\begin{deluxetable}
\begin{table} %[h]
\begin{center} 
\caption{Basic parameters assumed for the model}
\begin{tabular}{lll}
\tableline
\tableline
%\noalign{\smallskip}
Parameter & Symbol  & Value  \\
%\noalign{\smallskip}
\tableline
%\noalign{\smallskip}
Mass of the compact object & $M_{\rm c}$ & 1.4 $M_{\sun}$\\ 
Jet's injection point & $z_0$ & 50 $R_{\rm g}^{\;\;1}$  \\
%Opening angle of the jet ?& $\theta$ &5$^\circ$\\
Initial radius & $R_0$ & $z_0/10$ \\
Mass of the companion star & $M_\star$ & 10 $M_{\sun}$\\ 
Radius of the companion star & $R_\star$ &10 $R_{\odot}$ \\
Effective temperature of the star & $T_{\rm eff}$& 22500 K\\ 
%Mass loss rate & $\dot{M}_*$ & $10^{-5}$ $M_{\odot}$ yr$^{-1}$\\ 
Density of the wind at the base& $\rho_0$& $10^{-11}$ gr cm$^{-3}$\\
Initial wind velocity & $v_0$ & 5 km s$^{-1}$\\
%Compact object accretion rate & $\dot{M}_{\rm disk}$ & $10^{-8}$ $M_{\odot}$ yr$^{-1}$ \\ 
%Wind velocity index & $\beta$ & 1 \\
%Shock's efficiency  & $\chi$ & 0.1\\
%Jet's expansion index & $n$ & 2 \\
Jet's Lorentz factor & $\Gamma$ & 1.25 \\
Minimum proton energy & ${E'}_p^{\rm min}$ & 1 GeV \\
Maximum proton energy & ${E'}_p^{\rm max}$ & 100 TeV \\
Penetration factor & $f_{\rm p}$ & 0.1 \\
%Orbital axis & $a$ & 2 $R_*$\\
Eccentricity & $e$ & 0.72 \\
Orbital period & $P$ & 26.496 d\\
Index of the jet proton distribution &$\alpha$& 2.2\\
%\noalign{\smallskip}
\tableline \multicolumn{3}{l} {$^1$$R_{\rm g}=GM_{\rm c}/c^2$.}\cr
\end{tabular}
\end{center} 
\label{t1}
%\end{deluxetable}
\end{table}

\clearpage

\clearpage

\clearpage
\begin{figure}%[h]
\epsscale{.90}
\plotone{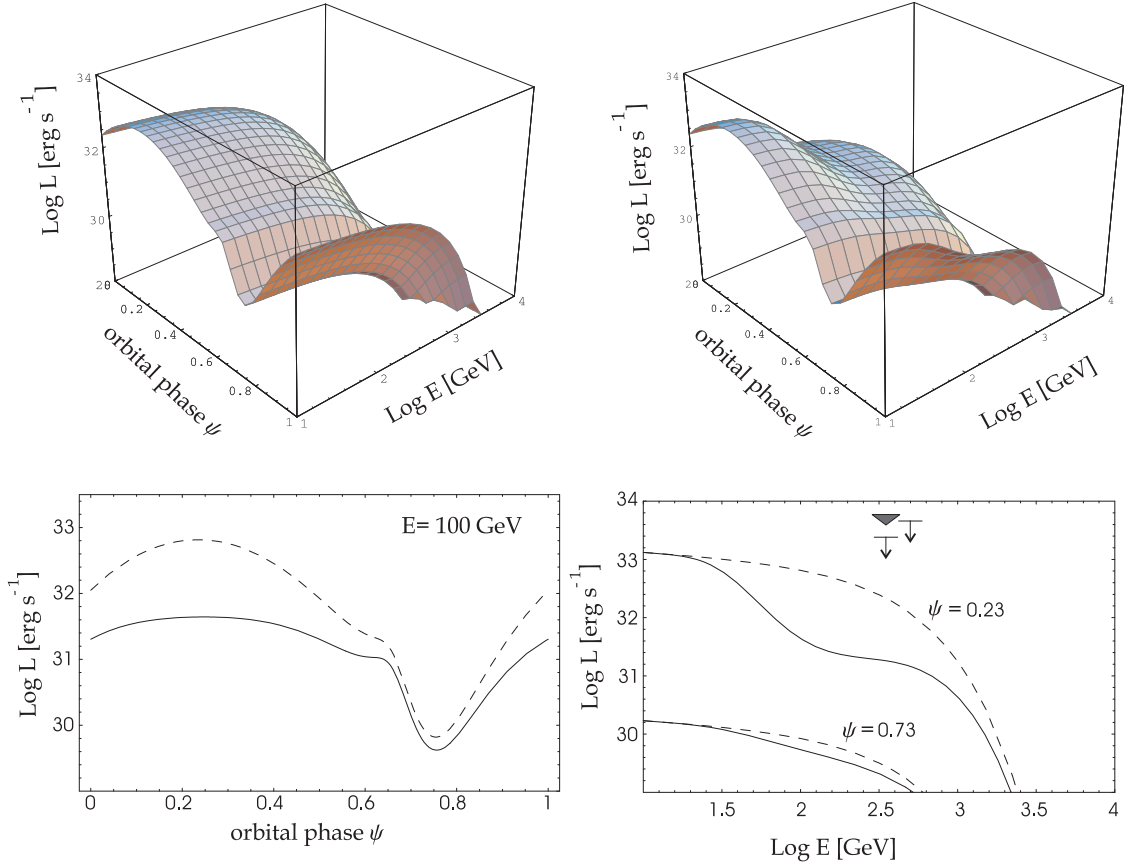} \caption{Upper, Left: A three-dimensional plot shows the generated luminosity as function of the orbital phase and gamma-ray energy. Right: The same plot, taken into account the gamma-ray absorption in the ambient photon fields.  Down, Left: Ligthcurve for gamma-rays of energy 100 GeV. The dashed curve corresponds to the generated luminosity, whereas the continuous curve takes into account the effects of photospheric opacity. Right: Spectral energy distribution at the periastron and apastron passage. The unabsorbed spectra are in dashed lines. Upper limits from Whipple observations are indicated.}\label{Lum}
\end{figure}

\clearpage

\clearpage

\end{document}